\documentclass{ws-ijmpa}

\begin{document}

\markboth{Shi-Lin Zhu}{Pentaquark}

\catchline{}{}{}{}{}

\title{Theoretical Development on Pentaquarks}

\author{Shi-Lin Zhu}
\address{Department of Physics, Peking University, BEIJING 100871, CHINA}


\maketitle


\begin{abstract}
I first give a brief overview of the status of exotic hadron
search so far. Then I review the recent theoretical development on
pentaquarks. Finally I emphasize that none of theoretical
formalisms predicts the existence of $\Theta$ pentaquark very
reliably. None of them explains its narrow width in a natural way.
\end{abstract}

\keywords{pentaquark, diquark}


\section{Status of exotic hadron search}

Quark model is very successful in the classification of hadron
states. Any state with quark content other than $q\bar q$, $q q q$
is beyond quark model. But quark model can't be the whole story.
Quantum chromodynamics as the underlying theory of strong
interaction allows much richer hadron spectrum such as: glueball,
hybrid meson/baryon, multiquark states, hadron molecules etc.

In the meson sector, there are several non-conventional meson
candidates. $f_0(980)/a_0(980)$ lies 10 MeV below the $K^+K^-$
threshold. It's difficult to find a suitable position for them
within the framework of quark model. So they were postulated to be
candidates of kaon molecule or four quark states. There are two
hybrid meson candidates with exotic quantum numbers
$J^{PC}=1^{-+}$: $\pi (1400), \pi (1600)$. However their masses
are much lower than theoretical predictions from various
approaches such as lattice QCD, QCD sum rule formalism and the
flux tube model. Scalar glueballs may mix strongly with nearby
scalar $q\bar q$ mesons. It's very demanding to establish the
glueball candidates $f_0(1500), f_0(1700)$ experimentally although
lattice QCD study indicates that pure scalar glueballs lie around
$1500$ MeV.

In the baryon sector, $\Lambda(1405)$ lies only 30 MeV below the
kaon nucleon threshold. It was postulated to be a kaon-nucleon
molecule. Because of its low mass, Roper resonance was suggested
to be a hybrid baryon with explicit glue inside. There does not
exist a hybrid baryon with exotic quantum numbers. According to
some models, the excitation of glue degree's of freedom carries an
energy around $0.8$ GeV which renders the hybrid baryon to be much
heavier than N(1440). Jaffe's H particle and other dibaryons
candidates carry two units of baryon numbers. Till the end of
2002, NONE of these non-conventional hadron states has been
established without controversy!

Since early 2003, there have been very important advances in
hadron spectroscopy. BES collaboration observed the $p\bar p$ and
$p \Lambda$ near-threshold enhancement in $J/\Psi$ decays
\cite{bes}. If these enhancements arise from a meson, they should
also be observed in the mesonic final states \cite{gao}. BABAR,
CLEO and BELLE collaborations observed two narrow charm-strange
mesons $D_{sJ}(2317), D_{sJ}(2457)$ below threshold \cite{babar}.
These states may admit a small portion of $D K$ or $D^\ast K$
continuum contribution in their wave functions. But the dominant
component of $D_{sJ}(2317), D_{sJ}(2457)$ should be $c\bar s$
\cite{dai}. BELLE collaboration discovered a new narrow charmonium
state X(3872) \cite{x}. Its production rate is comparable to that
of other excited charmonium states. Some people suggested that it
could be a deuteron-like $D {\bar D}^\ast$ molecule with four
constituent quarks. Then it's very difficult to explain its big
production rate \cite{chao}. Recently SELEX collaboration reported
a narrow state $D_{sJ}(2632)$ above threshold with abnormal decay
pattern \cite{selex}. Possible interpretations were suggested in
Refs. \cite{liu}. The most important event is LEPS and other
collaborations reported evidence of pentaquarks \cite{leps}. A
recent review can be found in Ref. \cite{ijmpa}.

\section{Pentaquarks: discovery or artifact}

Since last year LEPS and many other collaborationx reported the
$\Theta^+$ pentaquark at $1.54\pm 0.01$ GeV with a width smaller
than 25 MeV \cite{leps}. Such a resonance with $B=+1, S=+1$ and
the minimum quark content $uudd\bar s$ is clearly beyond the
conventional quark model. In the 2004 PDG $\Theta^+$ has been
listed as a three-star resonance.

However, many other experimental groups reported negative results
\cite{nega}. One may wonder how significant these negative
experiments are. Some of them didn't see $\Sigma^\ast(1670),
\Sigma^\ast(1660)$, which are four or three-star states in PDG!
Certainly we can not rule out the existence of them based on the
non-observation of these negative experiments. Moreover,
production mechanism could be different for pentaquarks at high
and low energy!

For those positive experiments, the following discrepancy should
be clarified. $\Theta$ mass extracted from $nK^+$ final states
differs significantly from that extracted from the $pK_s$ channel.
The width of $\Theta$ from direct measurements by HERMES and ZEUS
groups is around ten MeV, which is much larger than the upper
bound set by the indirect analysis of kaon nucleus scattering
data. Readers should keep in mind that the existence of $\Theta$
is still not completely established.

\section{Theoretical Developments on Pentaquarks}

If $\Theta$ really exists, there are serious theoretical
challenges in the understanding of its low mass, very narrow width
and weird production mechanism. In the following I list the main
results of some theoretical approaches. I will not cover topics
such as (1) non-clustered quark models; (2) pentaquark production;
(3) $\Theta$ relatives; (4) heavy pentaquarks etc.

\subsection{Lattice QCD }

Lattice QCD starts from the first principle and should be the most
reliable approach in the hadron spectroscopy. However the present
status of lattice QCD calculation of pentaquark mass and parity is
very confusing. I collect the available lattice results in the
following table. Four groups observed pentaquark-like signals with
negative parity on the lattice. One of them observed positive
parity $\Theta$ pentaquark.

\vspace{0.2cm}

\begin{tabular}{|c|c|c|c|}
  \hline
  Groups & Signal  & Parity & Operators \\
  \hline
  Csikor et al. \cite{csikor}     & Yes & Negative & Color-Octet Kaon Nucleon \\
  Sasaki \cite{sasaki}& Yes & Negative & Diquark-Diquark-Anti-quark \\
  Kentucky \cite{liukf}& No & Negative & Color-Singlet Kaon Nucleon \\
  Chiu-Hsieh \cite{chiu} & Yes & Positive & Diquark-Diquark-Anti-quark \\
  MIT \cite{mit}& Yes & Negative & Diquark-Diquark-Anti-quark \\
  Ishii et al. \cite{ishii} & No & Not Positive& Diquark-Diquark-Anti-quark \\
  Takahashi et al. \cite{taka} & Yes & Negative & Color-Singlet(-Octet) KN \\
  \hline
\end{tabular}

\subsection{Chiral soliton model and large $N_c$ expansion}

Diakanov et al. predicted $\Theta$ pentaquark at 1530 MeV in the
chiral soliton model (CSM) with $S=+1, B=1, J^P={1\over 2}^+, I=0,
\Gamma_\Theta< 15$ MeV \cite{dpp}. They assumed N(1700) as a
member of the anti-decuplet and used its mass and width as the
benchmark to fix $\Theta$ mass and width. Recently CSM was updated
by Ellis et al \cite{ellis}. With reasonable parameters, there
always exists a low-lying $\Theta$ pentaquark: $1430< m_\Theta
<1660$ MeV. The anti-decuplet spectrum can be reproduced with
bigger value of $\sigma$-term. The narrow width arises from the
cancellation of the coupling constant in the first three orders in
the large $N_c$ expansion.

But collective quantization may be invalid for pentaquarks in CSM
\cite{cohen1}. In this model, the mass splitting for non-exotic
excitation is ${\cal O}(1/N_c)$. For exotic excitaions, it is
${\cal O}(N_c^0)$. For pentaquarks, collective (rotation) and
vibration modes are not orthogonal. They mix at the leading order
of $N_c$ due to Wess-Zumino term. Collective quantization is NOT
legitimate for exotic motion in CSM. The successful prediction of
$\Theta$ properties in CSM is fortuitous \cite{cohen1}.

Large $N_c$ formalism does NOT predict $\Theta$ pentaquark.
However, given the existence of $\Theta$, one can predict its
large Nc partners in the same way that Delta is the $N_c$ partner
of nucleon \cite{cohen2}. The mass splitting between these states
is ${\cal O}(1/N_c)$. QCD consistency can be applied to scattering
amplitudes to derive relations between coupling constants etc.

\subsection{QCD sum rules}

Similarly, QCD sum rules (QSR) does NOT predict $\Theta$! Instead,
if it really exists, QSR can accommodate $\Theta$ at 1530 MeV and
yield the absolute mass scale without fit to data \cite{zsl}. All
QSR calculations \cite{zsl,oka,lee} receive continuum
contamination because of the high dimension of five-quark
interpolating currents. QSR with old-fashion perturbation approach
favors negative parity for $\Theta$ \cite{oka}. But the slow
convergence of OPE series in this case is a serious problem
\cite{comment}. Appropriate interpolating currents such as
color-octet KN type, diquark-diquark type can minimize the
coupling with the scattering states as indicated by lattice QCD.

\subsection{Diquark model}

Jaffe and Wilczek proposed the diquark model for pentaquarks
\cite{jaffe}. They assumed $\Theta$ is composed of one $\bar s$
and a pair of $J^P=0^+, {\bar 3}_F, {\bar 3}_c$ [ud] diquark.
Diquarks obey Bose statistics. Within their model, there exist an
antidecuplet and octet with $J^P={1\over 2}^+, {3\over 2}^+$. They
are nearly degenerate. After ideal mixing, the lowest state is
nucleon-like, which is identified as Roper resonance. But it is
hard to dynamically generate the scalar diquark with a low mass
420 MeV even with models. It's difficult to accommodate $\Theta$
and $\Theta_c(3099)$ simultaneously. Possible reasons could be (1)
the scalar diquark mass is too low; (2) anti-quark related
attraction is important and ignored in the diquark model. Some
lattice simulations reproduced N(1440) using three quark
interpolating currents, supporting Roper as nucleon's first radial
excitation. N(1440) has a width of 350 MeV while the upper bound
of $\Theta$ width is 1 MeV. It's really hard to accommodate them
simultaneously. Introduction of diquarks leads to many additional
low-mass states beyond quark model such as four-quark states
composed of a pair of diquark and anti-diquark. Where are they?

There exists $J^P={1\over 2}^-$ pentaquarks in diquark Model
\cite{zhang,wise}. When the flavor wave function of two diquarks
is anti-symmetric, one gets an octet and singlet with $L=0,
J^P={1\over 2}^-$. In other words, there are nine additional
pentaquark states lighter than $\Theta$, close to the orbitally
excited three-quark states. Especially two additional
$\Lambda$-like pentaquarks lie around $\Lambda(1405)$. $\Lambda
(1405)$ could be a pentaquark candidate! Two nucleon-like
pentaquarks may be narrow. The lowest heavy pentaquark has
$J^P={1\over 2}^-$ instead of $J^P={1\over 2}^+$. Experimental
discovery of these states is a direct support of the diquark
model. Otherwise, one should re-evaluate the relevance of this
picture.

The diquark and anti-quark are very similar in many respects.
There exists a supersymmetry between diquark and anti-quark
\cite{dib,lich}. If the diquark model is correct, pentaquarks can
be related to anti-baryons through replacing two diquarks by two
anti-quarks. Replacing the anti-quark by the diquark, pentaquarks
can be related to dibaryons composed of three diquarks. With the
experimental constraint on H dibaryon, one can estimate P=-
$SU_F$(3) singlet pentaquark mass to be around(1402-1542) MeV.
With experimental $\Theta$ mass, P=- dibaryon mass is found to be
around (2270-2310) MeV \cite{dib}.

\subsection{Other clustered quark models}

There are other clustered quark models. Karliner and Lipkin
proposed the diquark-triquark model \cite{lipkin}. Two clusters
were kept apart by the P-wave angular momentum barrier. Color
magnetic interaction occurs within two clusters only. Color
electric forces bind them into a color singlet with P=+. There
exist an anti-decuplet and octet. In order to get a low lying
$\Theta$, they assumed orbital excitation to be only 200 MeV to
set scale. If one uses the realistic value of the spin-average
orbital excitation around 500 MeV, $\Theta$ mass should be around
1800 MeV in this model as pionted out by Hogaasen and Sorba
\cite{hogaasen}!

Shuryak and Zahed suggested replacing one scalar diquark in JW's
model by a tensor one would lower $\Theta$ mass \cite{shuryak}.
There is one orbital excitation within the tensor diquark. No
orbital excitation between two diquarks is needed. But now
$\Theta$ is in the 27 representation! One may wonder where
$\Theta$'s isospin partners are.

\section{Narrow width puzzle}

Around $\Theta$ mass region, $\Lambda(1520)$ decays via D-wave
into K N. Its width is 7 MeV. $\Lambda(1600)$ decays via P-wave.
Its width is 100 MeV. Both decays need the $q\bar q$ pair creation
from vacuum. In contrast, $\Theta$ falls apart easily with similar
phase space. But its width is less than 10 MeV from direct
measurements and less than 1 MeV from indirect constraints. This
is a big puzzle.

There are some attempts for narrow width puzzle including: (1)
Color flavor spin overlap suppression \cite{jaffe,kim,close}; (2)
Spatial overlap suppression \cite{stech}; (3) Cancellation of
coupling constants in the first three orders in the large NC
expansion in CSM \cite{dpp,ellis}; (4) Decoupling from the decay
modes through the diagonalization of the mixing mass matrix
between two degenerate pentaquarks \cite{lipkin2}; (5) Special
diamond spatial structure \cite{song}.

\section{Summary}

If $\Theta$ is confirmed and established by future experiments, a
new landscape of multi-quark hadrons is emerging from the horizon!
We MUST answer: What's the underlying dynamics leading to its low
mass, narrow width and special production mechanism? Do other
multiquark hadrons exist: 4q,  6q, 7q, $\cdots$, N q? Is there an
upper limit for N? Study of these issues will deepen our
understanding of the low-energy sector of QCD!

At present, it is fair to say that NONE of theoretical models
predicts the existence of $\Theta(1530)$ reliably. NONE of
theoretical models naturally explains its low mass, narrow width
and production mechanism convincingly if it really exists. The
spin and parity of $\Theta$ distinguish theoretical models. Now
its existence is an experimental issue, requiring high-statistics
experiments. 

\section*{Acknowledgments}

This project was supported by the National Natural Science
Foundation of China under Grant 10375003, Ministry of Education of
China, FANEDD and SRF for ROCS, SEM.


\end{document}